\input{aipcheck}

\documentclass[
    ,final            
  ]
  {aipproc}
  
\layoutstyle{6x9}
\usepackage{amsmath,amssymb}

\def\ba{\begin{array}}
\def\ea{\end{array}}
\def\ovl{\overline}

\begin{document}

\title{Z$^{\bf \prime}$ Bosons from E$_{\bf 6}$:\\
        Collider and Electroweak Constraints}


\author{Jens Erler}{
  address={Departamento de F\'isica Te\'orica, Instituto de F\'isica, \\
                   Universidad Nacional Aut\'onoma de M\'exico, 04510 M\'exico D.F., M\'exico}
}

\author{Paul Langacker}{
  address={School of Natural Sciences, Institute for Advanced Study, \\
                   Einstein Drive, Princeton, NJ 08540, USA}
}

\author{Shoaib Munir}{
  address={Department of Physics, COMSATS Institute of Information Technology, \\
                   Defence Road, Lahore--54000, Pakistan}
}

\author{Eduardo Rojas}{
  address={Departamento de F\'isica Te\'orica, Instituto de F\'isica, \\
                   Universidad Nacional Aut\'onoma de M\'exico, 04510 M\'exico D.F., M\'exico}
}

\begin{abstract}
Many models beyond the Standard Electroweak Theory, top-down or bottom-up,  
contain extensions of the gauge symmetry group by extra $U(1)^\prime$ factors 
which can be understood or treated as subgroups of $E_6$.
A brief overview of such models is followed by a sketch of a systematic classification. 
We then describe how the resulting extra massive neutral gauge bosons can be searched
for and in case of positive evidence diagnosed using electroweak and collider data.
\end{abstract}

\maketitle


\section{Introduction}

$Z^\prime$ bosons~\cite{Langacker:2008yv} are among the best motivated kinds of physics beyond the Standard Model (SM). 
They easily appear in top-down scenarios like Grand Unified Theories or superstring constructions.
In fact, it often requires extra assumptions if one wants to avoid an additional $U(1)^\prime$ gauge symmetry 
or decouple the associated $Z^\prime$ from observation. This is even more true in bottom-up
approaches where $U(1)^\prime$ symmetries are a standard tool to alleviate problems in models of 
dynamical symmetry breaking, supersymmetry, large or warped extra dimensions, little Higgs, {\em etc}.
And as all these models are linked to electroweak symmetry breaking, the $Z^\prime$ mass, $M_{Z'}$,
should be in the TeV region, providing a rationale why they might be accessible at current or near future 
experiments. 

\begin{table}[t]
\begin{tabular}{ccrrccrr}
\hline
$l \equiv \left( \ba{c} \nu \\ e^- \ea \right)$ & & $-2 c_2$ & $-c_3$ & \hspace{95pt}
$\ba{c} \bar{\nu} \\ e^+ \ea$ & $\ba{r} -c_1 \\ +c_1 \ea$ & $\ba{r} +c_2 \\ +c_2 \ea$ & $\ba{r} +2 c_3 \\ +2 c_3 \ea$ \\
\hline
$q \equiv \left( \ba{c} \phantom{l} u \phantom{l} \\ \phantom{l} d \phantom{l} \ea \right)$ & & & $+c_3$ & \hspace{95pt}
$\ba{c} \bar{u} \\ \bar{d} \ea$ & $\ba{r} -c_1 \\ +c_1 \ea$ & $\ba{r} -c_2 \\ -c_2 \ea$ & \\
\hline\hline
$L \equiv \left( \ba{c} N \\ E^- \ea \right)$ & $-c_1$ & $+c_2$ & $-c_3$ & \hspace{95pt}
$\ba{c} D \\ \ovl{D} \ea$ & & $\ba{r} \\ +2 c_2 \ea$ & $\ba{r} -2 c_3 \\ {} \ea$ \\
\hline
$\ovl{L} \equiv \left( \ba{c} E^+ \\ \ovl{N} \ea \right)$ & $+c_1$ & $+c_2$ & $-c_3$&  \hspace{95pt}
$S$ & & $\ba{r} -2 c_2 \ea$ & $\ba{r} +2 c_3 \ea$ \\
\hline
\end{tabular}
\caption{Charge assignment for the left-handed multiplets contained in a $\bf{27}$.
The upper part corresponds to the $\bf{16}$ of $SO(10)$, 
while the lower part shows the $\bf{10}$ (with an extra anti-quark weak singlet, 
$\ovl{D}$, of electric charge $-1/3$ and an additional weak doublet, $L$,
as well as their SM-mirror partners) and the~$\bf{1}$ (a SM singlet, $S$).  
This represents one fermion generation, and family universality is assumed.}
\label{tab:e6charges}
\end{table}

$Z^{\bf \prime}$ discovery would most likely occur as an $s$-channel resonance at a collider, 
but interference with the photon or the standard $Z$ provides leverage also at lower energies.
Once discovered at a collider, angular distributions may give an indication of its spin to discriminate it
against states of spin 0 ({\em e.g.} the sneutrino of supersymmetry) and spin 2 (like the Kaluza-Klein graviton
in extra dimension models). 
The diagnostics of its charges would be of utmost importance as they can hint at the underlying principles.

\section{Z$^\prime$ bosons from E$_{\bf 6}$}

$E_6$ is the only exceptional compact Lie group that possesses chiral representations
and is therefore a candidate for unified model building. 
At the same time, all representations of $E_6$ are free of gauge anomalies
so that any $U(1)'$ subgroup of $E_6$ corresponds to a $Z'$ candidate,
independently of whether $E_6$ provides a successful unification group\footnote{$E_6$ symmetry generally 
predicts unacceptable Yukawa terms involving the additional "exotic" particles contained in its 
fundamental ${\bf 27}$ dimensional representation.}.
$Z'$ models with the same charges for the SM fermions also arise from a bottom-up approach~\cite{Erler:2000wu} 
when demanding the cancellation of anomalies in supersymmetric extensions of the SM 
together with a set of fairly general requirements such as allowing the SM Yukawa couplings,
gauge coupling unification, 
a solution~\cite{Suematsu:1994qm,Cvetic:1995rj} to the $\mu$-problem~\cite{Kim:1983dt}, 
the absence of dimension~4 proton decay as well as fractional electric charges,
and chirality (to protect all fields from acquiring very large masses). 
The most general such $Z'$ can be written as,
$$
Z' = \cos\alpha \cos\beta\, Z_\chi + \sin\alpha \cos\beta\,  Z_Y + \sin\beta\,  Z_\psi 
    = {c_1\, Z_R + \sqrt{3}\, (c_2\, Z_{R_1} + c_3\, Z_{L_1}) \over \sqrt{c_1^2 + 3\, (c_2^2 + c_3^2)}},
$$
where $-\pi/2 < \beta \leq \pi/2$ is the mixing angle between the $U(1)_\chi$ and $U(1)_\psi$ maximal subgroups 
defined by~\cite{Robinett:1982tq} $SO(10) \to SU(5) \times U(1)_\chi$ and $E_6 \to SO(10) \times U(1)_\psi$, respectively,
and $-\pi/2 < \alpha \leq \pi/2$ is non-vanishing when there is a mixing term~\cite{Holdom:1985ag} 
between the hypercharge, $Y$, and $U(1)'$ field strengths $\propto F_Y^{\mu\nu} F'_{\mu\nu}$,
and this kinetic mixing term has been undone by field redefinitions. 
The $U(1)_Y$, $U(1)_\chi$, and $U(1)_\psi$ groups are mutually orthogonal, and so are
the $U(1)_R$, $U(1)_{R_1}$, and $U(1)_{L_1}$, which are 
defined by $SU(3)_{L,R} \to SU(2)_{L,R} \times U(1)_{L_1,R_1}$ and $SU(2)_R \to U(1)_R$, referring here 
to the trinification subgroup~\cite{Achiman:1978vg} of $E_6 \to SU(3)_C \times SU(3)_L \times SU(3)_R$.
For the $U(1)'$ charges in terms of the $c_i$ we refer to Table~\ref{tab:e6charges} 
and for a complete discussion and classification to Ref.~\cite{ER:2011}.

\begin{figure}
  \includegraphics[height=.26\textheight]{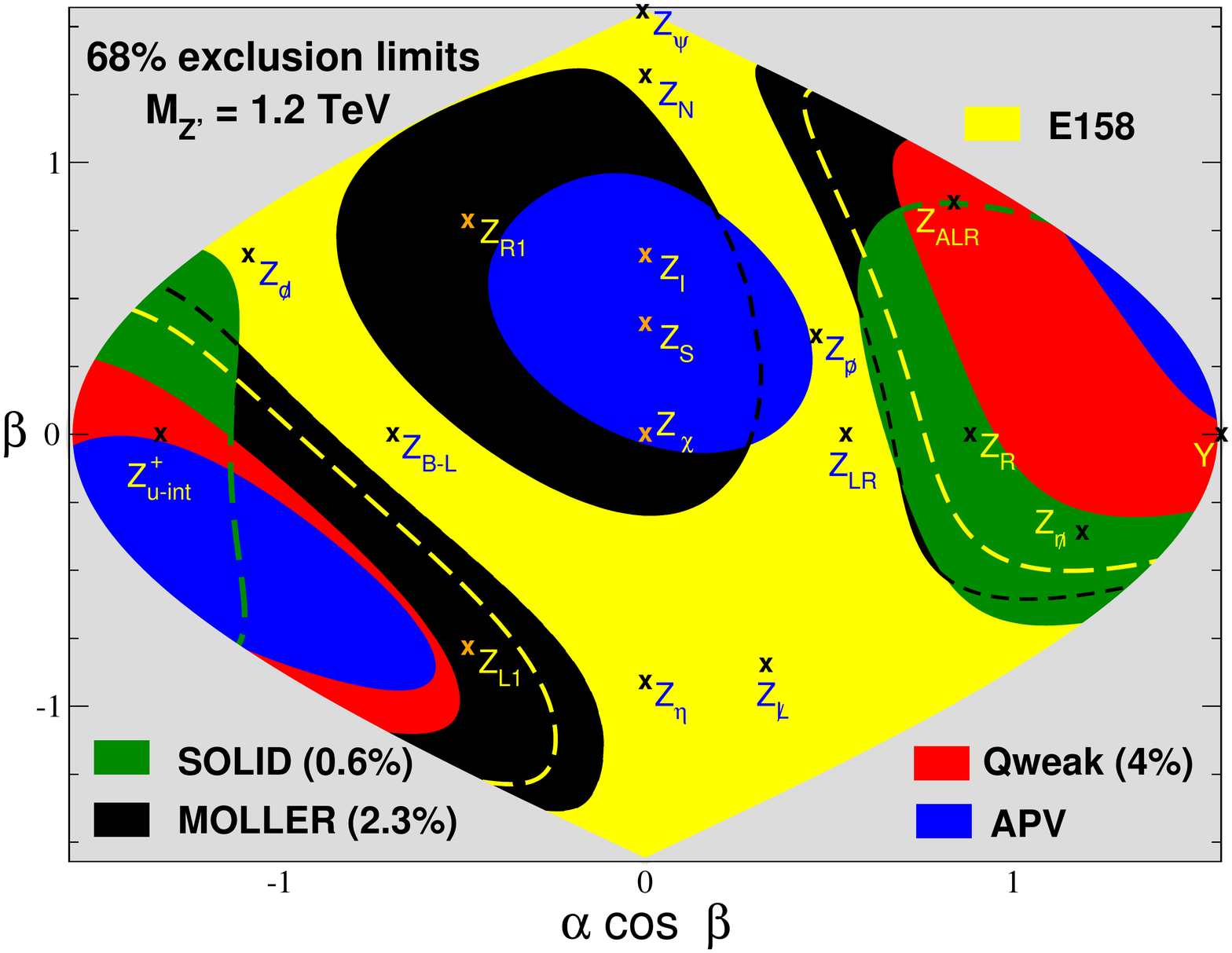}
  \includegraphics[height=.26\textheight]{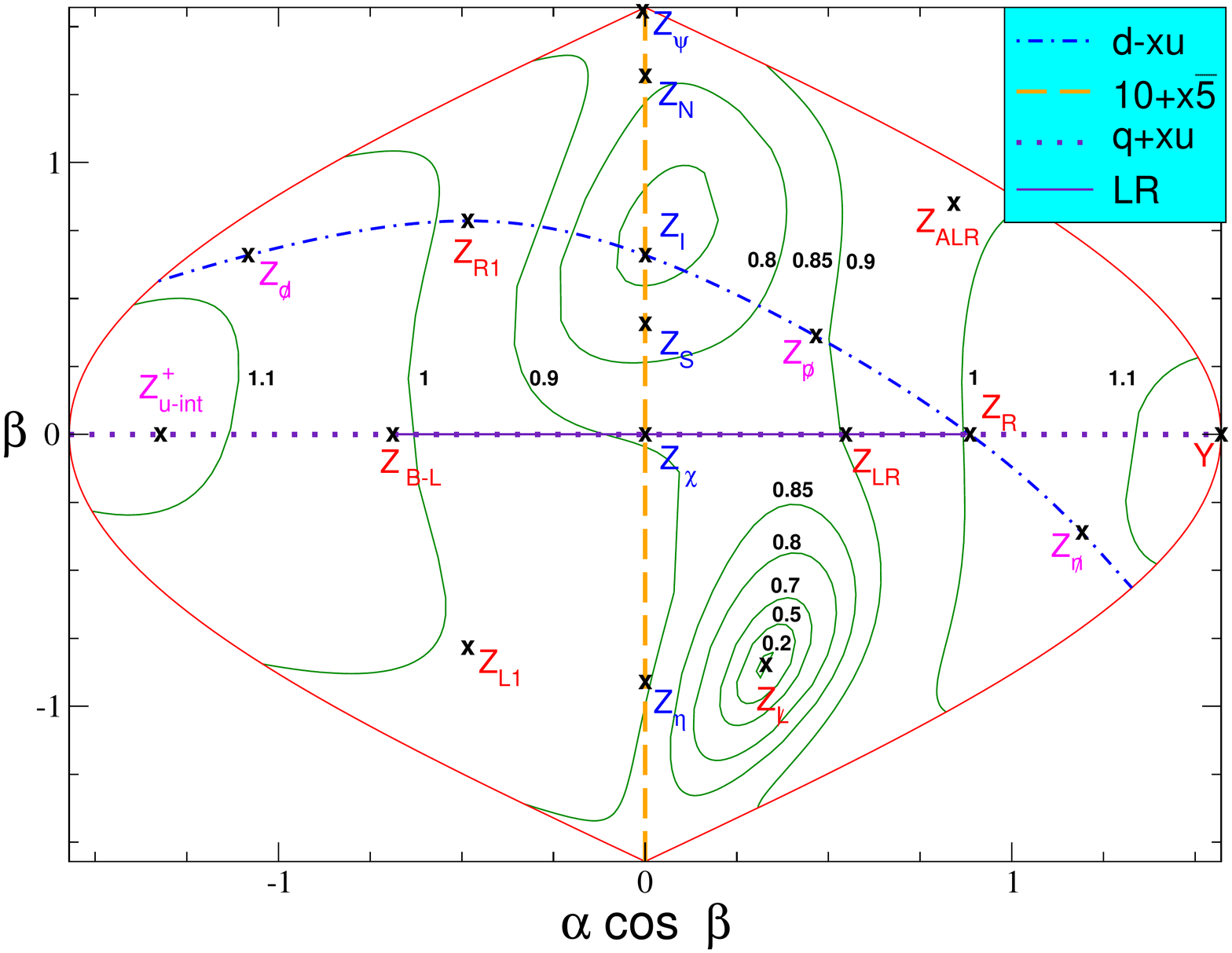}
  \caption{Left: 68\% CL exclusion constraints on $E_6$ inspired $Z'$ models provided by parity violating M\o ller 
  scattering (E158) and atomic parity violation (APV). 
  Projections for future JLab asymmetries from MOLLER ($ee$), Qweak (elastic $ep$), and SOLID (eDIS) are also shown. 
  Right: 95\% CL $M_{Z'}$ exclusion contours (in TeV) based on CDF di-muon data~\cite{Aaltonen:2008ah}.  
  (see Ref.~\cite{Erler:2011ud} for details).
  Notice the blind spot around the leptophobic $Z_{\not L}$ and the complementarity of these data sets. 
  The horizontal (purple) line indicates $SO(10)$ (including left-right) models.
  The vertical (yellow) line corresponds to vanishing kinetic mixing.  
  The dash-dotted (blue) line marks the $U(1)_{d-xu}$ model class~\cite{Carena:2004xs} corresponding to $c_3 = 0$.}
  \label{fig1}
\end{figure}

\section{Z$^\prime$ constraints from electroweak precision data}
The $Z$ pole experiments at LEP~1 and the SLC precisely determined the vector and axial-vector $Z$ couplings
and constrained the $ZZ'$ mixing angle to $\sin\theta_{ZZ'} \lesssim \mbox{few} \times 10^{-3}$.
While these experiments were virtually blind to $Z'$ bosons with negligible $ZZ'$ mixing,
precision measurements at much lower energies (away from the $Z$ pole) 
can probe the $Z'$ exchange amplitude {\em via\/} its interference with the photon
when the pure photon contribution is cancelled in polarization asymmetries. 
Furthermore, this class of experiments can be used to constrain 
the $Z'$ charges as illustrated in Fig.~\ref{fig1}.
Note that the combination of these experiments may cover the entire coupling space (for $M_{Z'} = 1.2$~TeV).

\begin{figure}
  \includegraphics[height=.42\textheight]{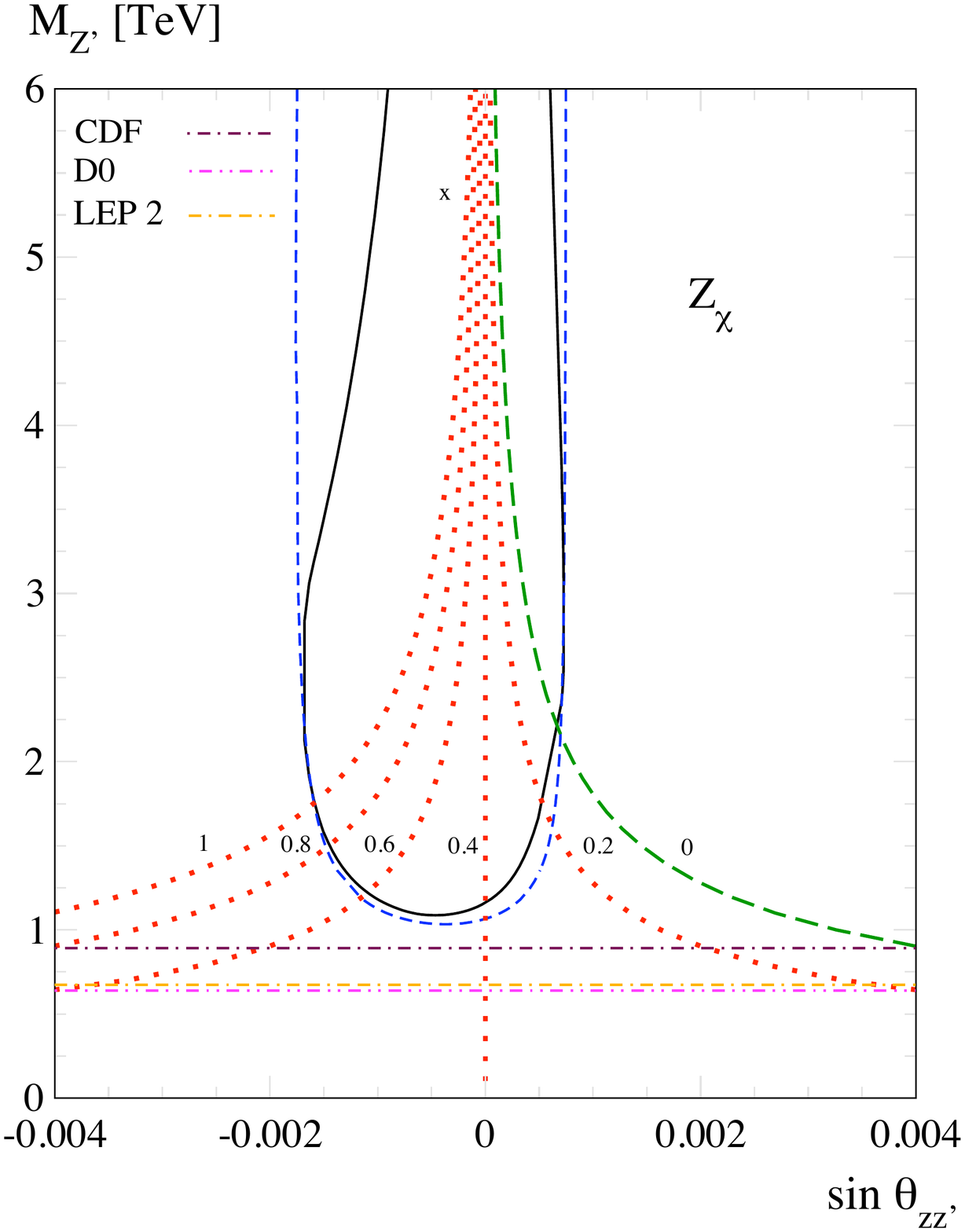}
  \includegraphics[height=.42\textheight]{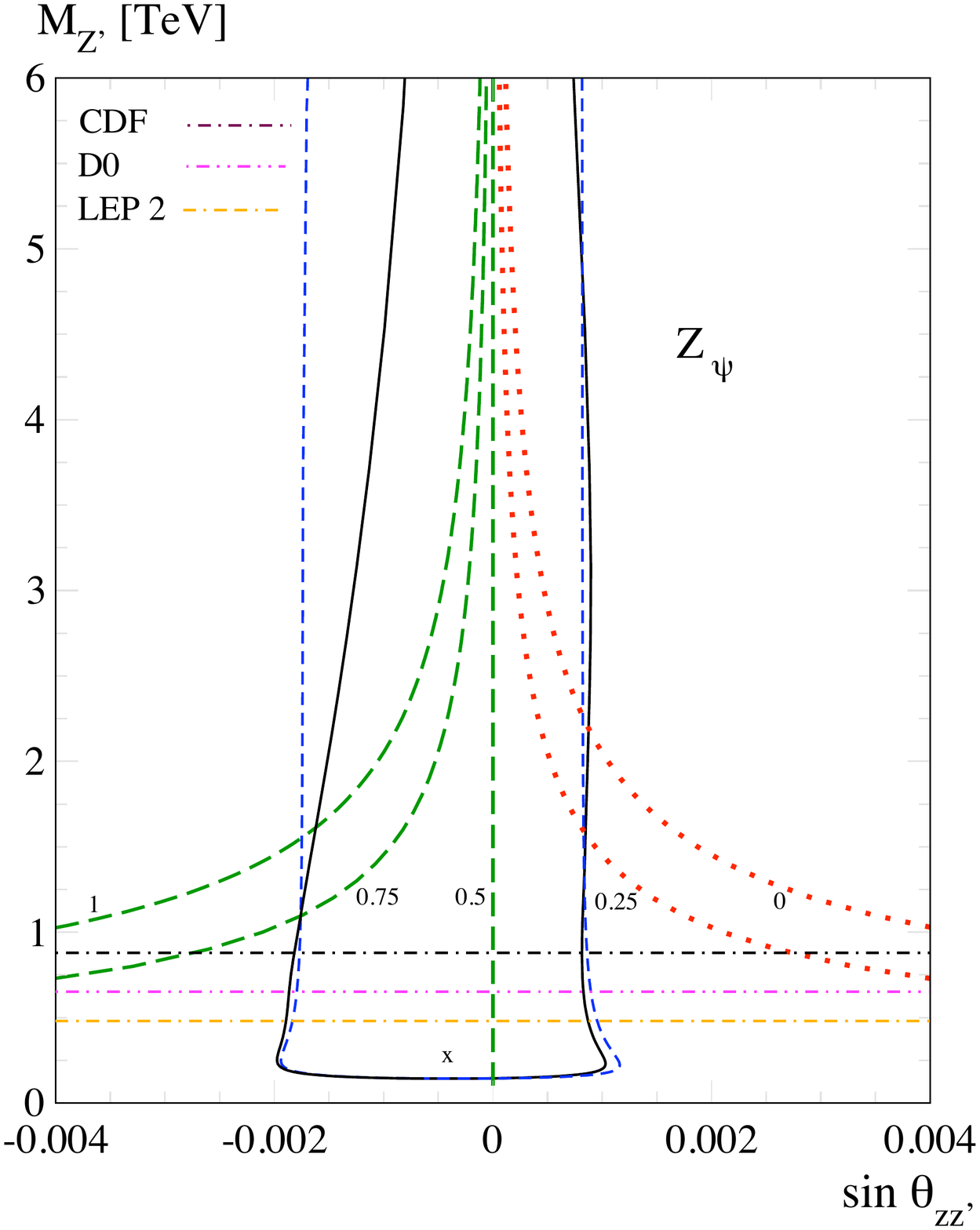}
  \caption{Precision constraints (95\% CL contours) on $M_{Z'}$ and $\sin\theta_{ZZ'}$. See Ref.~\cite{Erler:2009jh} for details.}
  \label{fig2}
\end{figure}

Fig.~\ref{fig2} shows the constraints from the combination of all precision data,
including the $Z$ lineshape, $Z$ pole asymmetries, the heavy flavor sector, the top quark and $W$ boson masses,
and from parity violation at low energy.
Also shown as dotted (red) and dashed (green) lines are the additional constraints arising for special (simple) Higgs sectors.
The latter usually arise in supersymmetric models when one wants to avoid the spontaneous breaking of 
lepton number and problems with charged current universality, as well as non-perturbative values for the top quark Yukawa coupling.
Thus, precision data can simultaneously constrain the $U(1)'$ breaking Higgs sector.
We also emphasize that the preferred range of the SM Higgs mass is often raised by the presence of the $Z'$~\cite{Erler:2009jh}.

\begin{figure}[b]
  \includegraphics[height=.26\textheight]{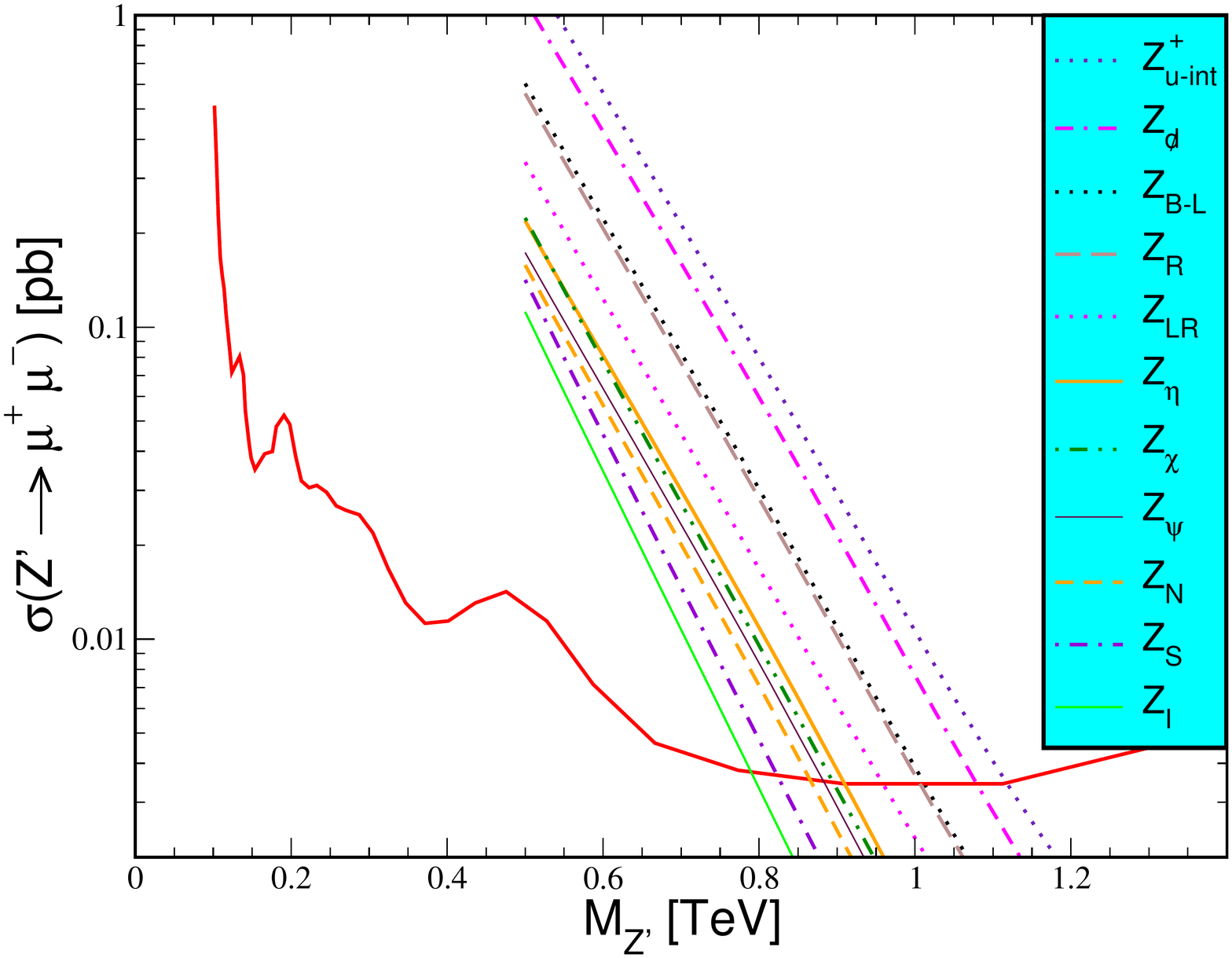}
  \includegraphics[height=.26\textheight]{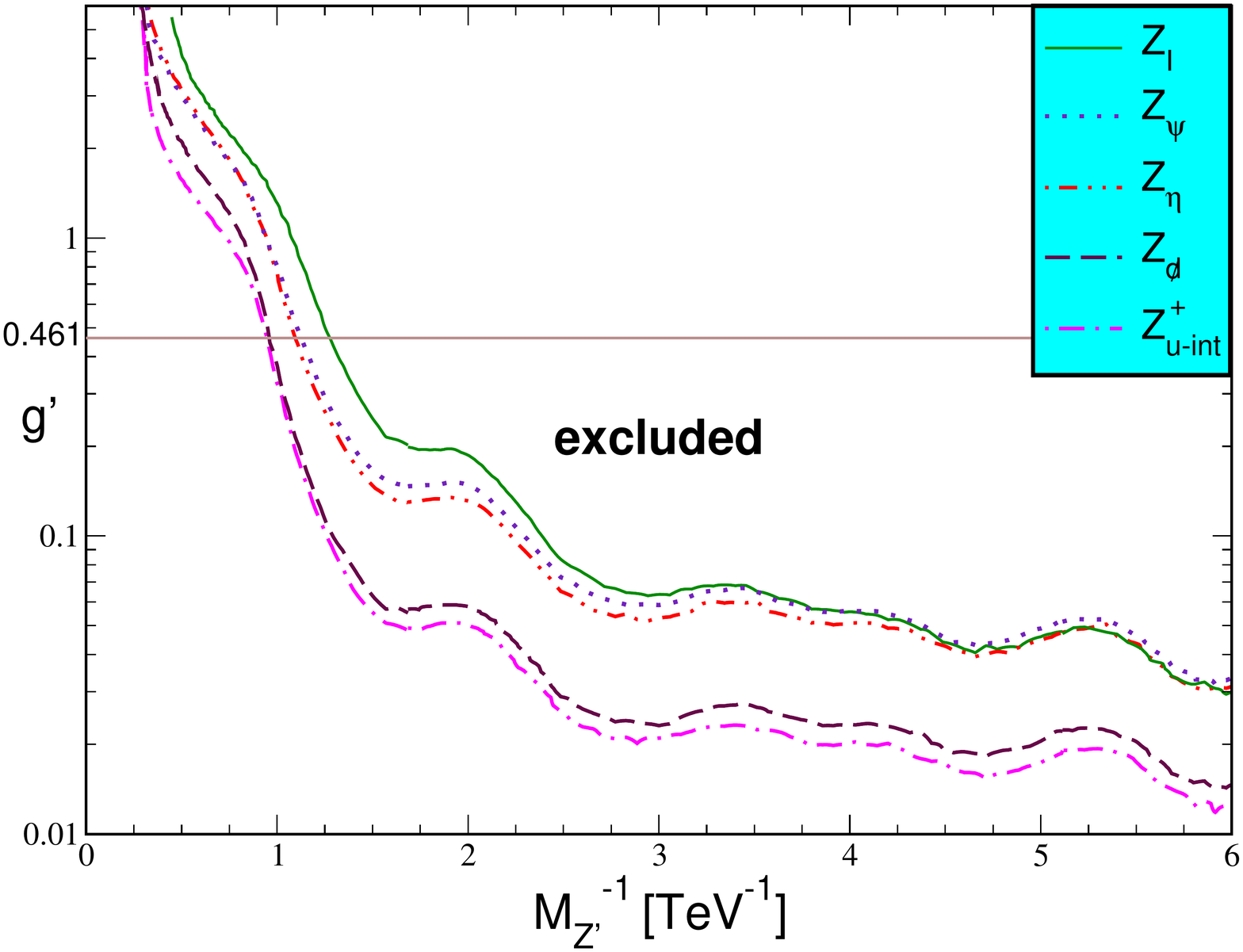}
  \caption{Left: classical (frequentist style) presentation of $Z'$ searches as used by CDF and D\O.  
  The intersection of the cross-section limit (solid line) with the model lines yields the $M_{Z'}$ bounds. 
  Right: $M_{Z'}$ limits as a function of $g'$. 
  While na\"ively, bump hunting would produce a straight vertical line, there must be non-trivial $M_{Z'}$ 
  dependence as can be seen by considering the limit $g' \rightarrow 0$ where the $Z'$ decouples.}
  \label{fig3}
\end{figure}

\section{Z$^\prime$ physics at hadron colliders}
One looks for $Z'$ bosons in resonant production of $e^+ e^-$,  $\mu^+ \mu^-$, $b\bar{b}$,  $t\bar{t}$, and di-jets
at hadron colliders. Forward-backward asymmetries serve as an additional diagnostic tool. 
Mass limits are obtained by comparing measured cross-sections with the predictions of
various $Z'$ models as a function of $M_{Z'}$ (see Fig~\ref{fig3}). More information can be gained, however,
from the ratio of likelihoods of signal plus background to the SM-only hypotheses as a function of $g'$,
with a mild dependence on parton distributions~\cite{Erler:2011ud} 
(we treated the mass dependent acceptance as a Beta distribution, and included QED and QCD effects).

\subsection{Towards an integrated Z$^\prime$ analysis}
\begin{figure}
  \includegraphics[height=.26\textheight]{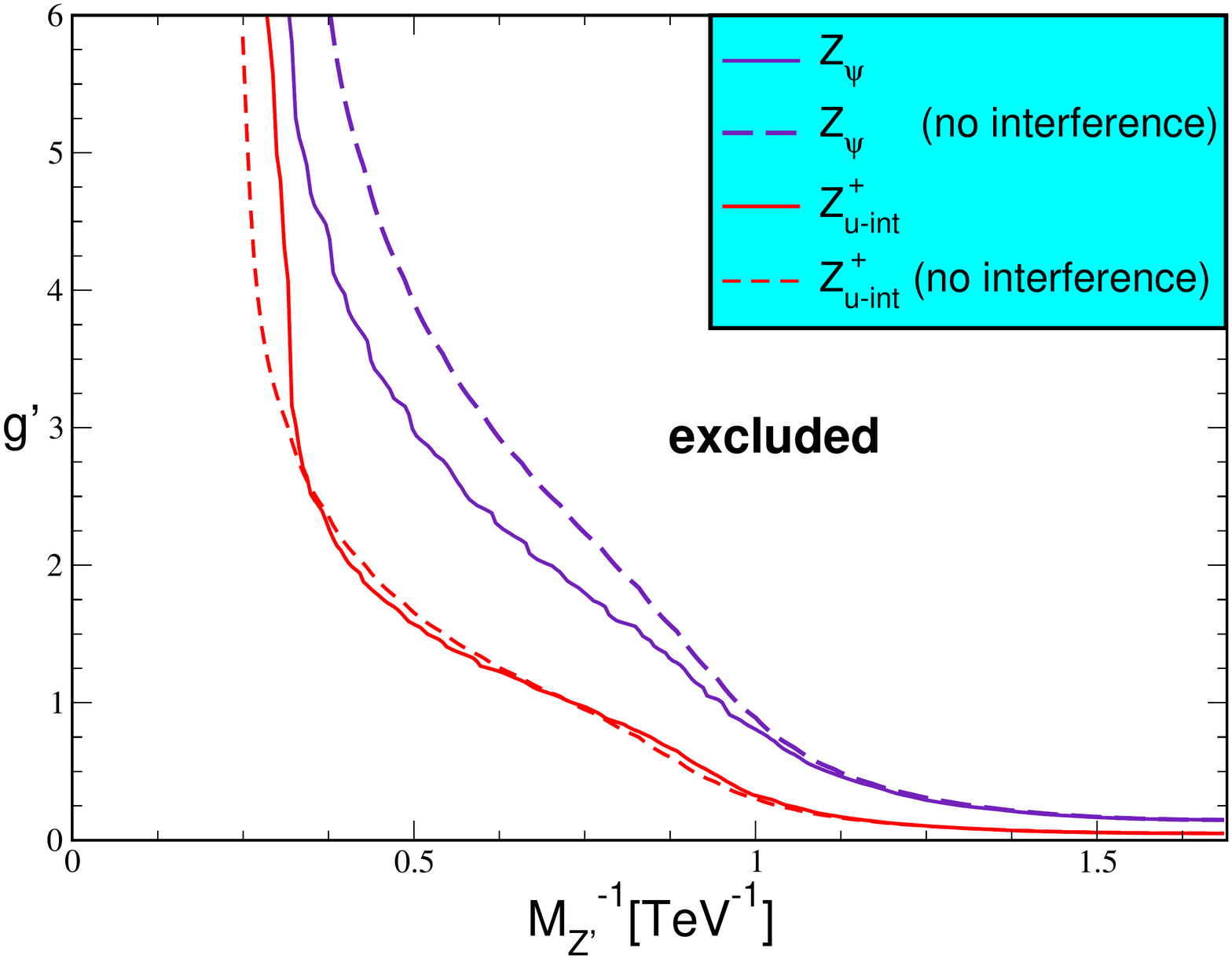}
  \includegraphics[height=.26\textheight]{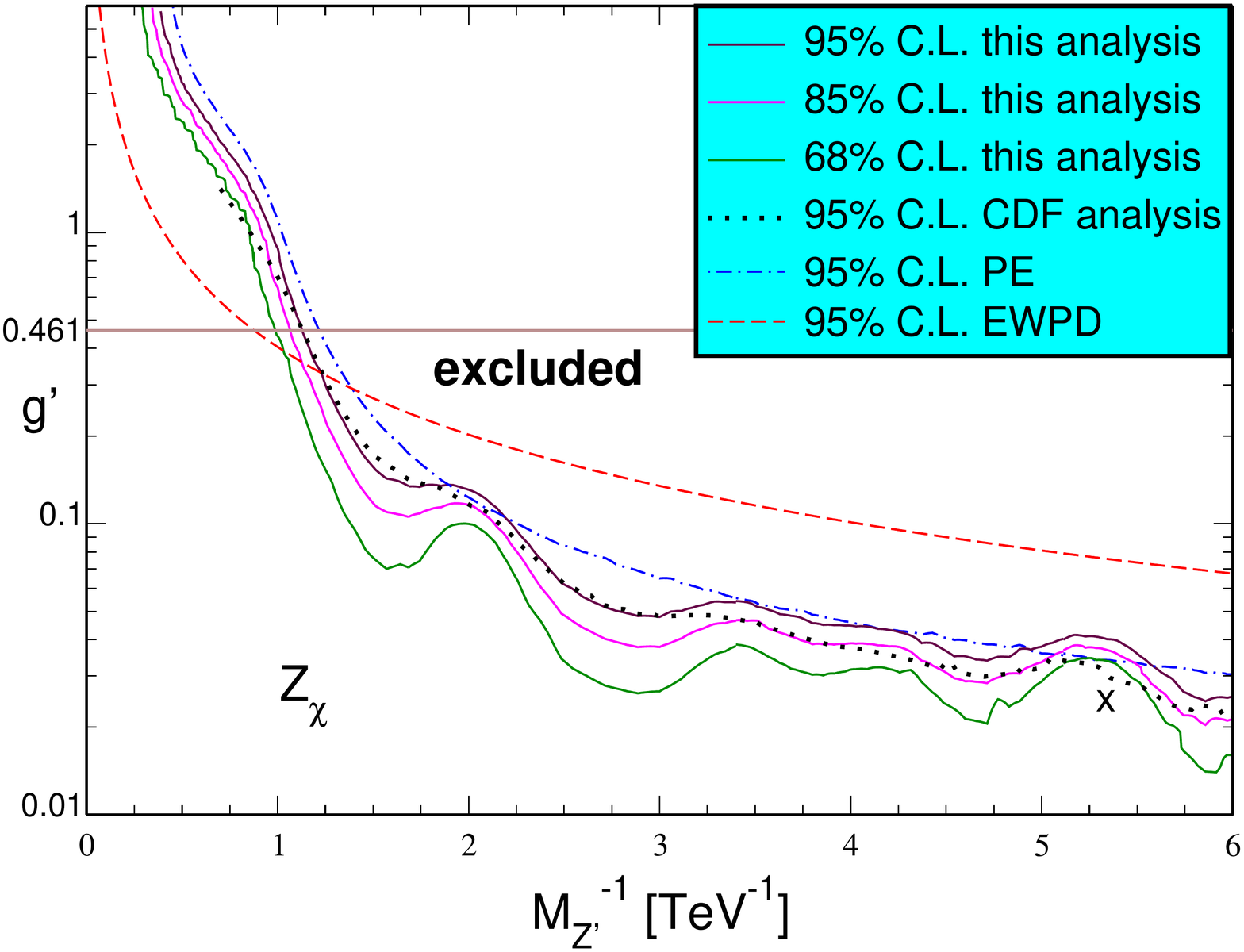}
  \caption{Left: interference effects are easily included in our approach,
  shown here for cases with enhanced constructive ($Z_{\rm u-int}^+$) and destructive ($Z_\psi$) interference.
  At large $M_{Z'}$, the pure $Z'$ contribution is suppressed with interferences dominating, 
  and it is seen that they can be significant even for moderate $M_{Z'}$ provided $g' \gtrsim 1$.
  We found that they can deepen the $\chi^2$ minimum, drastically change its location,
  and lift a possible $\chi^2$ degeneracy between different models. 
  Right: comparative analysis of the $Z_\chi$ showing the complementarity of electroweak precision data (EWPD)
  and the di-muon analyses.
  Comparing pseudo-experiments (PE) with the actual analyses shows that the bumps in 
  the collider curves are real.}
  \label{fig4}
\end{figure}

This Bayesian style analysis (see Fig.~\ref{fig4}) also suggests the next step, {\em i.e.,} to move from a collection of mass limits to 
an integrated analysis, especially after the arrival of a hint or a discovery.
The log-likelihood constructed~\cite{ER:2011} for the specific di-muon data set of CDF~\cite{Aaltonen:2008ah} 
can be combined with analogous functions (of $\alpha, \beta, \theta_{ZZ'}, g', M_{Z'}, \dots $) corresponding to other Tevatron channels, 
LEP 1, SLC, LEP 2, the LHC, CEBAF, APV, {\em etc.}, and used to disentangle these parameters.
This amounts to an intensive long-term project, and warrants close collaboration between theorists and experimentalists. 
   
\begin{theacknowledgments}
The work at IF-UNAM is supported by CONACyT project 82291--F. 
The work of P.L. is supported by an IBM Einstein Fellowship and by NSF grant PHY--0969448.
E.R. acknowledges financial support provided by DGAPA--UNAM.
\end{theacknowledgments}



\bibliographystyle{aipproc}  

\bibliography{myreferences2}

\IfFileExists{\jobname.bbl}{}
 {\typeout{}
  \typeout{******************************************}
  \typeout{** Please run "bibtex \jobname" to optain}
  \typeout{** the bibliography and then re-run LaTeX}
  \typeout{** twice to fix the references!}
  \typeout{******************************************}
  \typeout{}
 }

\end{document}